\documentclass[fleqn, 12pt] {article}
\usepackage{epsf}
\begin{document}
\title{Development of nonlinear two fluid interfacial structures by combined action of Rayleigh-Taylor, Kelvin-Helmholtz and Richtmyer-Meshkov instabilities:Oblique shock}
\author{M. R. Gupta\thanks{e-mail: mrgupta$_{-}$cps@yahoo.co.in}, Labakanta Mandal\thanks{e-mail: laba.kanta@yahoo.com}, Sourav Roy \thanks{e-mail: phy.sou82@gmail.com}, Rahul Banerjee\thanks{e-mail:
rbanerjee.math@gmail.com},Manoranjan Khan\thanks{e-mail: mkhan$_{-}$ju@yahoo.com} \\
Dept. of Instrumentation Science \& Centre for Plasma Studies
\\Jadavpur University, Kolkata-700032, India\\}
\date{}
\maketitle

\begin{abstract}
The nonlinear evolution of two fluid interfacial structures like
bubbles and spikes arising due to the combined action of
Rayleigh-Taylor and Kelvin-Helmholtz instability or due to that of
Richtmyer-Meshkov and Kelvin-Helmholtz instability resulting from
oblique shock is investigated. Using Layzer's model analytic
expressions for the asymptotic value of the combined growth rate
are obtained in both cases for spikes and bubbles. However, if the
overlying fluid is of lower density the interface perturbation
behaves in different ways. Depending on the magnitude of the
velocity shear associated with Kelvin-Helmholtz instability both
the bubble and spike amplitude may simultaneously grow
monotonically (instability) or oscillate with time or it may so
happen that while this spike steepens the bubble tends to
undulate. In case of an oblique shock which causes combined action
of Richtmyer-Meshkov instability arising due to the normal
component of the shock and Kelvin Helmholtz instability through
creation of velocity shear at the two fluid interface due to its
parallel component, the instability growth rate-instead of
behaving as $1/t$ as $t \rightarrow \infty$ for normal shock, tends
asymptotically to a spike peak height growth velocity $\sim
\sqrt{\frac{5(1+A_{T})}{16(1-A_{T})}(\Delta v)^2}$ where $\Delta v
$  is the velocity shear and $A_T$ is the Atwood number. Implication of
such result in connection with generation of spiky fluid jets in
astrophysical context is discussed.
\end{abstract}
\newpage

\section*{I. INTRODUCTION}
Rayleigh-Taylor (RTI) and Kelvin-Helmholtz (KHI) instabilities are
associated with the perturbation of the interface of two fluids of
different densities subject to the action of continuously acting
acceleration (with respect to time) and under the action of
velocity shear,respectively. The perturbation and the consequent
instability may also be induced by a shock generated impulsive
acceleration known as Richtmyer-Meshkov (RMI) instability. Such
interfacial hydrodynamic instabilities occur in a wide range of
physical phenomenon from those associated with problems on wave
generation by wind blowing over water surface to problems related
to Inertial Confinement Fusion (ICF) or astrophysical problems
like that of supernova explosion remnant which belong to the
domain of high energy density (HED) physics $^{\cite{rd06}}$. In
ICF experiment HED plasmas may be created due to multi kilo Joule
laser with a pressure $\sim$ Mbar. In ICF,in addition to RT and RM
instabilities nonspherical implosion generate shear flows; the
later is also formed when shocks pass through irregular fluid
interfaces. The KHI and shear flow effects in general are also of
practical importance in a number of HED system. They should be
considered in a multi shock implosion schemes for direct drive
capsule for ICF, since KHI may accelerate the growth of turbulent
mixing layer at the interface between the ablator and solid
deuterium-tritium nuclear fuel. In HED and astrophysical system,
it has been seen that structures driven by shear flow appear on
the high density spikes produced by R-T and R-M
instabilities$^{\cite{kk03}}$. They may develop in course of
evolution of these instabilities $^{\cite{br06}}-^{\cite{db07}}$
and cover enormous range of spatial scales from $10^{17}$cm for
jets from young stellar objects to $10^{24}$cm for jets from
quasars or active galactic nuclei$^{\cite{br06}}$. Examples are
suggested to be provided by pillars ("elephant trunk") of Eagle
Nebula which are identified with spikes of a heavy fluid
penetrating a light fluid$^{\cite{ls54}},^{\cite{ef54}}$. Another
example in astrophysics is the Herbig-Haro (HH) object like HH34,
where jets are observed with knots. Buhrke et al $^{\cite{tb88}}$
explained that Kelvin-Helmholtz instability is the reason for
knots in the jets. The jet must be $\sim 10$ times denser than its
surrounding medium having velocity $\sim $300 km/sec and Mach no.
30. Steady isolated jets may form structure through the growth of
K-H modes. The stability properties of super magnetosonic
astrophysical jets are subject of current interest.

The linear theory of the combined effects of RT,KH and RM
instabilities have been investigated earlier
$^{\cite{m94}}$. Weakly nonlinear theoretical results of
Kelvin-Helmholtz and Rayleigh-Taylor instability growth rates
together with different aspects of density and shear velocity
gradients have also been discussed$^{\cite{lw09}}-^{\cite{lw10}}$.

In case of the temporal evolution of these instabilities nonlinear
structures develop at the two fluid interface. The structure is
called a bubble if the lighter fluid pushes across the unperturbed
surface into the heavier fluid and a spike if the opposite takes
place. The dynamics of such RTI and RMI generated nonlinear
structures have been studied $^{\cite{jh94}}-^{\cite{ps03}}$ under
different physical situation using an expression near the tip of
the bubble or spike up to second order in the transverse
coordinate to unperturbed surface following Layzer's
$^{\cite{mr09}}$approach.

In the present paper, we investigate the combined effect of
Rayleigh-Taylor,Richtmyer-Meshkov and Kelvin Helmholtz
instabilities by extending the above method so as to include the
effect of velocity shear induced contribution to the growth rate
of the tip of the nonlinear mushroom like structures generated by
shock wave (normal or oblique) incident on the unperturbed
interface.

In the event of excitation of RM instability due to normal
incidence of shock in absence of velocity shear of the growth rate
of the height of the finger like structures decay as
$(1/t)$$^{\cite{vn02}},^{\cite{ps03}}$. It is however interesting
to note that if the shock incidence is oblique (or if it passes
across an irregular surface) the growth rate of the tip of the
spiky structure does not decrease as $(1/t)$ but attains a
saturation value proportional to $\sqrt{k^2(
 \Delta v)^2/(1-A_{T})}$ where $\Delta v$=difference is the tangential
velocity of the fluids at the interface and $A_{T}$ is the Atwood
number. Thus the growth rate may be quite large if
$A_{T}\rightarrow 1$ which may be likely in astrophysical
situation and thus play an important role in formation of
jets$^{\cite{br06}} ,^{\cite{dd02}}$.

The paper is organized in the following manner. In section II is
developed the basic equations describing the dynamics of nonlinear
structures which evolve in consequence of the combined effects of
these different types of hydrodynamical instabilities. In section
III it is shown that the classical results$^{\cite{dl55}}$ follow
on linearization of the evolution equation describing the bubbles
and spikes. Numerical as well as some analytical results regarding
the saturation growth rates are presented in section IV. Finally
section V presents a brief summary of this results.

\section*{II. BASIC EQUATIONS OF EVOLUTION OF THE HYDRODYNAMIC INSTABILITIES}

Let the $y=0$ plane denote the unperturbed surface of separation
of two fluids (the line $y=0$ in the two dimensional form of this
problem). The fluid with density $\rho_{a}$ is assumed to overlie
the fluid with density $\rho_{b}$. The gravity
$\overrightarrow{g}$ is assumed to act along the negative y- axis.
Any perturbation of the horizontal interface or a shock driven
impulse gives rise to Rayleigh-Taylor instability($\rho_{a}>
\rho_{b}$) or Richtmyer -Meshkov instability which in course of
temporal evolution gives rise to nonlinear interfacial structures.

The two fluids separated by the horizontal boundary are further
assumed to be in relative horizontal motion and thus subjected to
Kelvin-Helmholtz instability arising due to horizontal velocity
shear. Thus we are faced with the problem of the combined action
of Rayleigh-Taylor and Kelvin-Helmholtz instabilities.We shall see
the same formulation will be applicable to Richtmyer-Meshkov
instability associated with an oblique shock  incident on the two
fluid interface.\\After perturbation the finger shaped interface
is assumed to take up a parabolic shape given by
\begin{eqnarray} \label{eq:1}
y(x,t)=\eta(x,t)=\eta_0(t)+\eta_2(t)(x-\eta_1(t))^2
\end{eqnarray}
For a bubble (here the lower fluid is pushing across the interface
into the upper fluid with density $\rho_{a}>$ density $\rho_{b}$)
we have,
\begin{eqnarray}\label{eq:2}
\mbox{}\qquad\qquad \eta_0>0 \quad \mbox{and} \quad \eta_2<0
\end{eqnarray}
and for spike:
\begin{eqnarray}\label{eq:3}
\qquad\qquad \eta_0<0 \quad \mbox{and} \quad \eta_2>0
\end{eqnarray}

The height of the vertex of the parabola i.e, the height of the
peak of the bubble (or spike) above the x-axis is $|\eta_{0}(t)|$.
The position of the peak at time t is at $x=\eta_{1}(t)$ and
because of the relative streaming motion of the two fluids the
peak moves parallel to the x-axis with velocity
$\dot{\eta}_{1}(t)$. The densities of both fluids are uniform and
fluid motion is supposed to be single mode potential flow.

For the upper fluid with density $\rho_{a}$ we take the velocity
potential
\begin{eqnarray}\label{eq:4}
\phi_a(x,y,t)=\left[\alpha_a(t)\cos{(k(x-\eta_{1}(t)))}+\beta_a(t)\sin{(k(x-\eta_{1}(t)))}\right]e^{-k(y-\eta_0(t))}-xu_{a}(t);
\quad y>0
\end{eqnarray}
and for the lower fluid (density $\rho_{b}$) the velocity
potential
\begin{eqnarray}\label{eq:5}
\phi_b(x,y,t)=\left[\alpha_b(t)\cos{(k(x-\eta_{1}(t)))}+\beta_b(t)\sin{(k(x-\eta_{1}(t)))}\right]e^{k(y-\eta_0(t))}-xu_{b}(t)+yb_{0}(t);
\quad y<0
\end{eqnarray}

Before proceeding with the analysis of the kinematic and boundary
conditions using the two fluid interface perturbation
$y=\eta(x,t)$ we forward the following justification for
restricting the expansion to terms O$(x-\eta_1(t))^2$. We are
concerned only motion very close to the tip of the bubble or spike
i.e., only in the region $k\mid x-\eta_1(t)\mid <<1.$ Consequently
one is justified in neglecting terms O $(\mid x-\eta_1(t) \mid
)^4$ unless the coefficients of such terms are sufficiently large.
Further it has been shown $^{\cite{mr10}}$ that even it terms
$\sim \eta_4(t)(x-\eta_1(t))^4 +\eta_6(t) (x-\eta_1(t))^6 $ are
retained the contribution from coefficients $\mid \eta_4 \mid ,
\mid \eta_6 \mid <<$ that from $\mid \eta_2 \mid $ at least in the
asymptotic state $\tau \rightarrow \infty$.

The kinematical boundary conditions satisfied at the interfacial
surface $y=\eta(x,t)$are
\begin{eqnarray}\label{eq:6}
\frac{\partial \eta}{\partial t}-\frac{\partial\phi_{a}}{\partial
x}\frac{\partial \eta}{\partial x}=-\frac{\partial \phi
_{a}}{\partial y}
\end{eqnarray}
\begin{eqnarray}\label{eq:7}
-\frac{\partial \phi_{a}}{\partial x}\frac{\partial \eta}{\partial
x}+\frac{\partial\phi_{a}}{\partial y}=-\frac{\partial
\phi_{b}}{\partial x}\frac{\partial \eta}{\partial
x}+\frac{\partial\phi_{b}}{\partial y}
\end{eqnarray}

The dynamical boundary conditions are next obtained from
Bernoulli's equation for the two fluids
\begin{eqnarray}\label{eq:8}
-\frac{\partial \phi_a}{\partial t}+ \frac{1}{2}(\vec{\nabla}
\phi_a)^2+gy\rho_a=-p_{a}+f_{a}(t)
\end{eqnarray}
\begin{eqnarray}\label{eq:9}
-\frac{\partial \phi_b}{\partial t}+ \frac{1}{2}(\vec{\nabla}
\phi_b)^2+gy\rho_b=-p_{b}+f_{b}(t)
\end{eqnarray}
by using the surface pressure equality
\begin{eqnarray}\label{eq:10}
p_{a}=p_{b}
\end{eqnarray}
leading to
\begin{eqnarray}\label{eq:11}
\nonumber-(\frac{\partial\phi_a}{\partial t}-\frac{\partial
\phi_b}{\partial t})+ \frac{1}{2}(\vec{\nabla} \phi_a)^2-
\frac{1}{2}(\vec{\nabla}
\phi_b)^2+g(\rho_a-\rho_b)y\\
=f_a(t)-f_b(t)
\end{eqnarray}
satisfied at the interface $y=\eta(x,t)$ Now from Eq.(1)
\begin{eqnarray}\label{eq:12}
\frac{\partial \eta}{\partial
t}=\dot{\eta}_0(t)-2\dot{\eta}_1(t)\eta_{2}(t)(x-\eta_{1}(t))+\dot{\eta}_2(t)(x-\eta_{1}(t))^2
\end{eqnarray}
Also utilizing the property that close to the tip of the bubble or
spike, $k|x-\eta_{1}(t)|<<1$, we express the velocity components
in the following form
\begin{eqnarray}\label{eq:13}
v_{ax}=-\frac{ \partial \phi_{a}}{\partial
x}=(u_{a}-k\beta_{a})+k^2\alpha_{a}(x-\eta_{1})+\beta
_{a}k^2(\eta_{2}+k/2)(x-\eta_{1})^2
\end{eqnarray}
\begin{eqnarray}\label{eq:14}
v_{ay}=-\frac{ \partial \phi_{a}}{\partial
y}=k\alpha_{a}+k^2\beta_{a}(x-\eta_{1})-k^2\alpha_{a}(\eta_{2}+k/2)(x-\eta_{1})^2
\end{eqnarray}
and similar expressions for $v_{bx}$ and $v_{by}$.

Following Layzer's$^{\cite{dl55}}$ model we substitute for
$\eta_{t},\eta_{x},(v_{a{(b)}})_{x},(v_{a{(b)}})_{y}$in the
kinematic and boundary conditions represented by Eqs.(6),(7)and
(11)and equate coefficients of $(x-\eta_{1}(t))^i;(i=0,1,2)$ and
neglect terms $O((x-\eta_{1}(t))^i);(i\geq 3)$.This yields the
following three algebraic equations for the three unknown
$b_{0},\alpha_{b},\beta_{b}$\,\,:
\begin{eqnarray}\label{eq:15}
b_0=-\frac{6\eta_2}{(3\eta_2-k/2)}k\alpha_a
\end{eqnarray}
\begin{eqnarray}\label{eq:16}
\alpha_b=\frac{(3\eta_2+k/2)}{(3\xi_2-k/2)}\alpha_a
\end{eqnarray}
\begin{eqnarray}\label{eq:17}
\beta_b=\frac{(\eta_2+k/2)k\beta_{a}-{\eta_2(u_{a}-u_{b})}}{k(\eta_2-k/2)}
\end{eqnarray}
and regarding the five other unknowns,viz
$\eta_{0}(t),\eta_{1}(t),\eta_{2}(t),\alpha_{a}(t),\beta_{b}(t)$
the following five nonlinear ODE's [Eqs.(18)-(22)].
\begin{eqnarray}\label{eq:18}
\frac{d\xi_1}{d \tau}=\xi_3
\end{eqnarray}
\begin{eqnarray}\label{eq:19}
\frac{d\xi_2}{d \tau}=-\frac{1}{2}(6\xi_2+1)\xi_3
\end{eqnarray}
\begin{eqnarray}\label{eq:20}
\nonumber
\frac{d\xi_3}{d\tau}=\frac{N_{1}(\xi_2,r)}{D_1(\xi_2,r)}\frac{\xi_3^2}{(6\xi_2-1)}+\frac{2(1-r)\xi_2(6\xi_2-1)}{D_1(\xi_2,r)}+\frac{N_{2}(\xi_{2},r)}{D_{1}(\xi_2,r)}\frac{(6\xi_2-1)\xi_4^2}{2\xi_2(2\xi_{2}-1)^2}
\hskip
70pt\\+2\frac{(4\xi_{2}-1)(6\xi_{2}-1)}{D_{1}(\xi_{2},r)(2\xi_{2}-1)^2}\left[(V_{a}-V_{b})^2\xi_2-(V_{a}-V_{b})(2\xi_{2}+1)\xi_{4}\right]
\end{eqnarray}
\begin{eqnarray}\label{eq:21}
\nonumber\frac{d\xi_4}
{d\tau}=\frac{(2\xi_2-1)}{D_2(\xi_2,r)}\left[(f_b-rf_a)-r\frac{\xi_3\xi_4}{2\xi_2}\right]+\frac{2(f_a-f_b)}{D_2(\xi_2,r)}\xi_2
\hskip
200pt\\
+\frac{(6\xi_2+1)\xi_3}{2D_2(\xi_2,r)(6\xi_2-1)(2\xi_2-1)}\left[4(V_a-V_b)(4\xi_2-1)-\frac{\xi_4}{\xi_2}(28\xi_2^2-4\xi_2-1)\right]
\end{eqnarray}
\begin{eqnarray}\label{eq:22}
\frac{d\xi_5}{d \tau}=V_a-\frac{\xi_4(2\xi_2+1)}{2\xi_2}
\end{eqnarray}
where \begin{eqnarray}\label{eq:23}
\xi_1=k\eta_0;\xi_2=\eta_2/k;\xi_5=k\eta_1\end{eqnarray}
\begin{eqnarray}\label{24}
\xi_3=k^2\alpha_a/\sqrt{kg};\xi_4=k^2\beta_a/\sqrt{kg},\tau=t\sqrt(kg)
\end{eqnarray}
\begin{eqnarray}\label{eq:25}
V_a=u_a\sqrt(k/g);V_b=u_b\sqrt(k/g);f_a=\frac{dV_a}{d
\tau};f_b=\frac{dV_b}{d \tau}.
\end{eqnarray}
The functions $N_{1,2}(\xi_2,r),D_{1,2}(\xi_2,r)$ where
$r=\frac{\rho_a}{\rho_b}$ is the density ratio are given by
\begin{eqnarray}\label{eq:26}
N_1(\xi_2,r)=36(1-r)\xi_{2}^{2}+12(4+r)\xi_{2}+(7-r)
\end{eqnarray}
\begin{eqnarray}\label{eq:27}
D_1(\xi_2,r)=12(r-1)\xi_{2}^{2}+4(r-1)\xi_{2}-(r+1)
\end{eqnarray}
\begin{eqnarray}\label{eq:28}
N_2(\xi_2,r)=16(1-r)\xi_2^3+12(1+r)\xi_2^2-(1+r)
\end{eqnarray}
\begin{eqnarray}\label{eq:29}
D_2(\xi_2,r)=2(1-r)\xi_2+(1+r)
\end{eqnarray}

The above set of five  Eqs. (18)-(22) together with Eqs.
(23)-(29)which define the different variables and functions
describe the combined effect of RT and KH instabilities.

On the other hand the impingement of an oblique shock on the two
fluid interface causes the joint effect of Richtmyer-Meshkov and
Kelvin-Helmholtz instability. The impact gives rise to an
instantaneous acceleration which will change the normal velocity
(y-component) by an amount $\Delta v=v_{after}-v_{before}$ and
transverse velocity (x-component) by $\Delta
u_{a(b)}=(u_{a(b)})_{after}-(u_{a(b)})_{before}$. Taking nonzero
values only for the post shock velocities we replace the
acceleration by their impulsive values. We set:
\begin{eqnarray}\label{eq:30}
\frac{d u_{a(b)}}{d t}=u_a\delta(t) \rightarrow\Delta v(t)
\end{eqnarray}
and replace $g\rightarrow \Delta v \delta(t)$ The dynamical
variables are non dimensionalized using normalization in terms of
$(k\Delta v)$instead of $\sqrt {kg}$.

The combined effect of RM-KH instability resulting from oblique
incidence of shock on the two fluid interface is then described by
the same set of equations as Eqs.(18)-(22) together with the
following replacements: \\$(i)$The second term on the RHS of
Eq.(20)drops out.\\(ii) $\xi_3,\xi_4$ and $\tau$ to be replaced by
$\overline{\xi_3}=\alpha_ak^2/(k\Delta
v),\overline{\xi_4}=\beta_ak^2/(k\Delta v)$ and $  \tau=t(k \Delta
v)$ respectively.\\(iii) $V_a$ and $V_b$ by $\overline V_a
=u_a/\Delta v,\overline V_b=u_b/\Delta v$.\\
(iv) $f_a$ by \begin{eqnarray}\label{eq:31} \overline f_a=\frac{d
\overline {v}_a}{d \overline {\tau}}=\frac{u_a}{\Delta v}\Delta
(\overline \tau) \quad \mbox{and} \quad f_b \quad \mbox{by} \quad
\overline{f_b}=\frac{d \overline {v}_b}{d \overline
{\tau}}=\frac{u_b}{\Delta v}\Delta (\overline \tau)
\end{eqnarray}

\section*{III. LINEAR APPROXIMATION}

We now show that the usual combined RT and KH instability growth
rates $^{\cite{sc81}}$ are recovered on linearization of Eqs.
(18)-(22). Let us put

\begin{eqnarray}\label{eq:31}
\nonumber\frac{d (k\eta_1)}{d
\tau}=\frac{d\xi_5}{d\tau}=\alpha_aV_a+\alpha_bV_b\,;\,\,\,\,
\left(\alpha_{a,(b)}=\frac{\rho_{a,(b)}}{\rho_{a}+\rho_{b}}\right)
\end{eqnarray}
in Eq.(22)giving
\begin{eqnarray}\label{eq:32}
\xi_4=2\alpha_b(V_a-V_b)\frac{\xi_2}{2\xi_2+1}\approx
2\alpha_b(V_a-V_b)\xi_2
\end{eqnarray}
on linearization . In absence of velocity shear $V_a-V_b=0$,we get
$\xi_4=0$. Thus the problem reduces to that of RT instability
alone with no contribution from KH instability. Linearizing Eqs.
(19),(20)and (21) we get
\begin{eqnarray}\label{eq:33}
\frac{d\xi_2}{d \tau}=-\frac{1}{2}\xi_3
\end{eqnarray}
\begin{eqnarray}
\frac{d\xi_3}{d\tau}=-2\left[A_T+\alpha_a\alpha_b(V_a-V_b)^2\right]\xi_2
\end{eqnarray}
\begin{eqnarray}\label{eq:34}
\frac{d\xi_4}{d \tau}=-\rho_a(V_a-V_b)\xi_3
\end{eqnarray}
$A_T=\frac{\rho_a-\rho_b}{\rho_a+\rho_b}$ is the Atwood number.
Eq.(32) connecting $\xi_2$ and $\xi_4$ provides the consistency
condition. The exponential growth rate due to combined effect of
RT and KH instability coincides with the classical linear theory
result $^{\cite{sc81}}$
\begin{eqnarray}\label{eq:35}\gamma (k)=\sqrt
{kg\left[A_T+\alpha_a\alpha_b(V_a-V_b)^2\right]}
\end{eqnarray}

\section*{IV. RESULTS AND DISCUSSIONS}
(A) \textbf{Combined effect of RT and KH instability}:

The growth rate of the RT instability induced nonlinear
interfacial structures is further enhanced due to KH instability.
Setting $\frac{d u_a}{dt}=0$ and $\frac{d u_a}{d t}=0$ the growth
rate of the peak height of the bubbles and spikes are obtained by
numerical integration of Eqs. (18)-(22) and the results are shown
in Fig.1. The dependence of the growth rate on $V_a$ and $V_b$
keeping $(V_a-V_b)$ unchanged are also indicated in the same
diagrams.It is found that for $V_b>V_a$ the growth rate is greater
than that for $V_a>V_b(\mid V_a-V_b \mid$ is the same for both
cases); the asymptotic values is the two cases are however
identical. Moreover for $\frac{\rho_a}{\rho_b}>1$ Eqs.(18)-(22)
show that as $\tau \rightarrow \infty$ there occurs growth rate
saturation given by
\begin{eqnarray}\label{eq:36}
(\xi_{3})_{bubble}^{asym}=\sqrt{\frac{2A_T}{3(1+A_T)}+\frac{5(1-A_T)}{16(1+A_T)}(V_a-V_b)^2}
\end{eqnarray}
and
\begin{eqnarray}\label{eq:37}
(\xi_{3})_{spike}^{asym}=\sqrt{\frac{2A_T}{3(1-A_T)}+\frac{5(1+A_T)}{16(1-A_T)}(V_a-V_b)^2}
\end{eqnarray}\
while
\begin{eqnarray}\label{eq:38}
\nonumber(\xi_{4})^{asym}=0
\end{eqnarray}
for both bubble and spike respectively.

Thus both saturation growth rate are enhanced for due to further
destabilization caused by the velocity shear.

On the other hand if $r=\rho_a/\rho_b<1
(A_T=\frac{\rho_a-\rho_b}{\rho_a+\rho_b}<0)$ there is no RT
instability but it follows from Eqs.(37)and (38) that instability
due to velocity shear (Kelvin -Helmholtz instability) persists on
both the wind ward side and leeward side (i.e; both for bubbles
and spikes) if (see Fig.2)
\begin{eqnarray}\label{eq:39}
\frac{32|A_T|}{15(1+A_T)}=\frac{16(1-r)}{15r}<(V_a-V_b)^2
\end{eqnarray}
and stabilized on both sides if (see Fig.3 which shows oscillation
of $\xi_1$and $\xi_3$ with respect to $\tau$)
\begin{eqnarray}\label{eq:40}
(V_a-V_b)^2<\frac{16}{15}(1-r)
\end{eqnarray}

If however $(V_a-V_b)^2$ lies in the interval specified by the
above inequalities,i.e;
\begin{eqnarray}\label{eq:41}
\frac{16}{15}(1-r)<(V_a-V_b)^2<\frac{16(1-r)}{15r};\mbox(r<1)
\end{eqnarray}
it follows from the same two Eqs.(37)-(38) that the peak of the
spike continues to steeper (instability) with $\tau$ as the
heavier fluid (density $\rho_b$)pushes across the interface into
the lighter fluid (density $\rho_a$) while the bubble height will
execute low finite amplitude undulations. The above observation is
shown to be suppressed in Fig.4. At time t, the peak height which
of the spike or the bubble occurs at $x=\eta_1(\tau)$ and thus
moves to the right (x-increases) as $\eta_1(\tau)$ increases with
$\tau$. The spike peak height increase monotonically with t while
that of the bubble undulates with low amplitude. The three
dimension representation of the steepening of the peak of the
spike as it moves along x-direction with time is shown in Fig.5.In
this respect there exists approximate qualitative agreement exists
with the results of the weakly nonlinear analysis$^{\cite{lw10}}$.

(B)  \textbf{Combined effect of Richtmyer-Meshkov and
Kelvin-Helmholtz instability: oblique shock}

The time evolution of the two fluid interfacial structure
resulting from the combined effect of Richtmyer-Meshkov and
Kelvin-Helmholtz instabilities consequent to impingement of an
shock is described by the set of Eqs.(18)-(22),(26)-(29) with
modifications as shown in the set of Eq.(31). If the shock
incidence is oblique then the normal component generates velocity
shear and causes KH instability.$^{\cite{m94}}$ The shock
generated initial values of $\overline \xi_3$ and $\overline
\xi_4$ are obtained from the impulsive accelerations represented
by the $ \delta -$ function terms in Eq.(30) giving
\begin{eqnarray}\label{eq:42}
(\overline
\xi_{3})_{\tau=0}=\left[\frac{2(1-r)\xi_2(6\xi_2-1)}{D_1(\xi_2,r)}\right]_{{(\xi_2)}{\tau=0}}
\end{eqnarray}
\begin{eqnarray}\label{eq:43}
(\overline
\xi_{4})_{\tau=0}=\frac{1}{D_2(\xi_2,r)}\left[\frac{(2\xi_2-1)(u_b-ru_a)+2\xi_2(u_a-u_b)}{\Delta
v}\right]_{(\xi_2)_{\tau=0}}
\end{eqnarray}

Results obtained from numerical solution of Eqs.(18)-(22) with
modifications given by Eq.(31) subject to initial conditions (42)
and (43) are presented in Fig.6. The growth rate contributed in
absence of velocity shear,i.e; by normally incident shock induced
Richtmyer-Meshkov instability varies as $t\rightarrow \infty$.
However in presence of velocity shear the growth rate due to
combined influence of RM and KH instability the growth rate
approaches finite saturation value asymptotically.

For RM-KH instability induced spikes it is given by the following
closed expression

\begin{eqnarray}\label{eq:44}
(\overline{\xi}_{3})_{t \rightarrow \infty}^{spike}=(\frac{d
\xi_1}{d t})_{t \rightarrow
\infty}^{spike}=\sqrt{\frac{5(1+A_T)}{16(1-A_T)}(u_a-u_b)^2/(\Delta
v)^2 }
\end{eqnarray}

which becomes large as the Atwood number $A_T \rightarrow $1
(equivalently$ \rho_a/\rho_b>>1)$

The following discussions suggest a higher plausibility of the
effectiveness of the joint influence of RM and KH instability in
the explanation of certain astrophysical phenomena.

Corresponding to parameter values for Eagle
Nebula($\rho_a/\rho_b=0.5\times10^2$ and $|u_a-u_b|=2\times10^6$
cm sec$^{-1}$)$^{\cite{br06}},^{\cite{dd02}}$ the velocity of rise
of the spike peak height h$^{spike}$(the height of the pillar)
according to Eq.(44)is $(\frac{d h}{d t})_{t \rightarrow
\infty}^{spike}\approx 0.79 \times 10^7$cm sec$^{-1}$.
Modification through inclusion of Rayleigh-Taylor instability
effect (see Eq.(38)) can only slightly increases this value to
$\approx 10^7$cm sec$^{-1}$.This gives the time to reach the
observed pillar height of $3\times 10^{19}$ cm of the Eagle Nebula
$\approx 10^4$ years. There are different time scales involved in
the problem of development of the pillar of the Eagle Nebula. As
pointed out by Pound, $^{\cite{mw98}}$ there is a characteristic
time scale for hydrodynamic motion $\tau_{dyn}\approx(\Delta
v)^{-1}$ where $\Delta v$ is the velocity shear inside the cloud.
Corresponding to data given in ref.(3) this turns out to be $\tau
_{hydrodanmic}\approx 10^{5}$yrs which is the upper time limit for
development  of the Eagle Nebula pillar("elephant trunk"). But it
is at least two orders of magnitude greater than the time scale
$\tau_{cool}\sim 10^2-10^3$ yrs imposed due to radiative cooling
of the cloud $^{\cite{br06}},^{\cite{dd02}}$. In comparison the
time scale of the development of the pillar is found here $\approx
10^4$yrs. Thus consequent to the hydrodynamic model based on the
combined influence of Richtmyer-Meshkov and Kelvin-Helmholtz
instability the gap between the two time scales $\tau_{cool}$ and
$\tau_{hydrodynamic}$is reduced by one order of magnitude.

A high Mach number, radiatevily cooled jet of astrophysical
interest has been produced in laboratory using intense laser
irradiation of a gold cone$^{\cite{df99}}$. The evolution of the
jet was imaged in emission and radiography.

K-H instability growth rate has recently been observed in HED
plasma experiment using Omega laser ($\lambda$)=0.351$\mu m$
delivering 4.3 $\pm 0.1$kJ to the target overlapping 10 drive
beams on to the ablator $^{\cite{eh09}}$.Incompressible K-H growth
rate peak to valley at Foam-Plastic interface has been compared
with several analytical modes.

\section*{V. Summary}
Finally we summarize the results:

(a)If the heavier fluid overlies the lighter fluid the growth rate
of both the bubble and spike peak heights due to RT instability
are enhanced due to concurrent presence of velocity shear, i.e,
K.H instability Fig.1. The asymptotic growth rates are given by
Eqs. (37) and (38).

(b) In the opposite case,i.e, if the overlying fluid is lighter
and lower one is heavier ($r=\rho_{a}/\rho_{b}<1$) both the spike
and bubble peak displacement increases continuously with time if
$\frac{16(1-r)}{15r}<(V_{a}-V_{b})^2$,i.e, instability persists
(Fig.3) while stabilization occurs if
$(V_{a}-V_{b})^2<\frac{(16(1-r)}{15})$ (Fig. 4 shows oscillation
of peak heights of bubbles and spikes).

(c) For $\frac{16(1-r)}{15}<(V_a-V_b)^2<\frac{16(1-r)}{15r}$ for
r$<1$ the spike steepens with time (the peak height continuously
increases with time as indicated in Fig.5. gives a three
dimensional graph of displacement y against x and $\tau$). But the
peak displacement of the bubble undulates within a small range
Fig.4.

(d)If the two fluid interface is subjected to an oblique shock
Kelvin-Helmholtz instability due to generation of velocity shear
occurs simultaneously with Richtmyer-Meshkov instability.The
growth rates of bubbles and spikes due to this joint action are
shown in Fig.6. respectively. It is important to note that the
growth rate of the combined action tends asymptotically to a
saturation value given by Eq.(44); this is in contrast to that due
to generation of RM instability due to normal shock incidence for
which the growth rate behaves as $\frac{1}{t}$ as t$ \rightarrow
\infty$. Moreover this growth rate as shown by Eq.(44) the rate of
growth of this spike height has sufficiently large magnitude if
the Atwood number $A_T \rightarrow $1($\rho_a<<\rho_b$).This may
have interesting implication in the hydrodynamic explanation of
formation of sufficiently long spiky jets in astrophysical
situation, e.g, in case of the Eagle Nebula.

\section * {ACKNOWLEDGEMENTS}
This work is supported by the Department of Science \& Technology,
Government of India under grant no. SR/S2/HEP-007/2008.

\begin{figure}[p]
\vbox{\hskip 1.5cm \epsfxsize=12cm \epsfbox{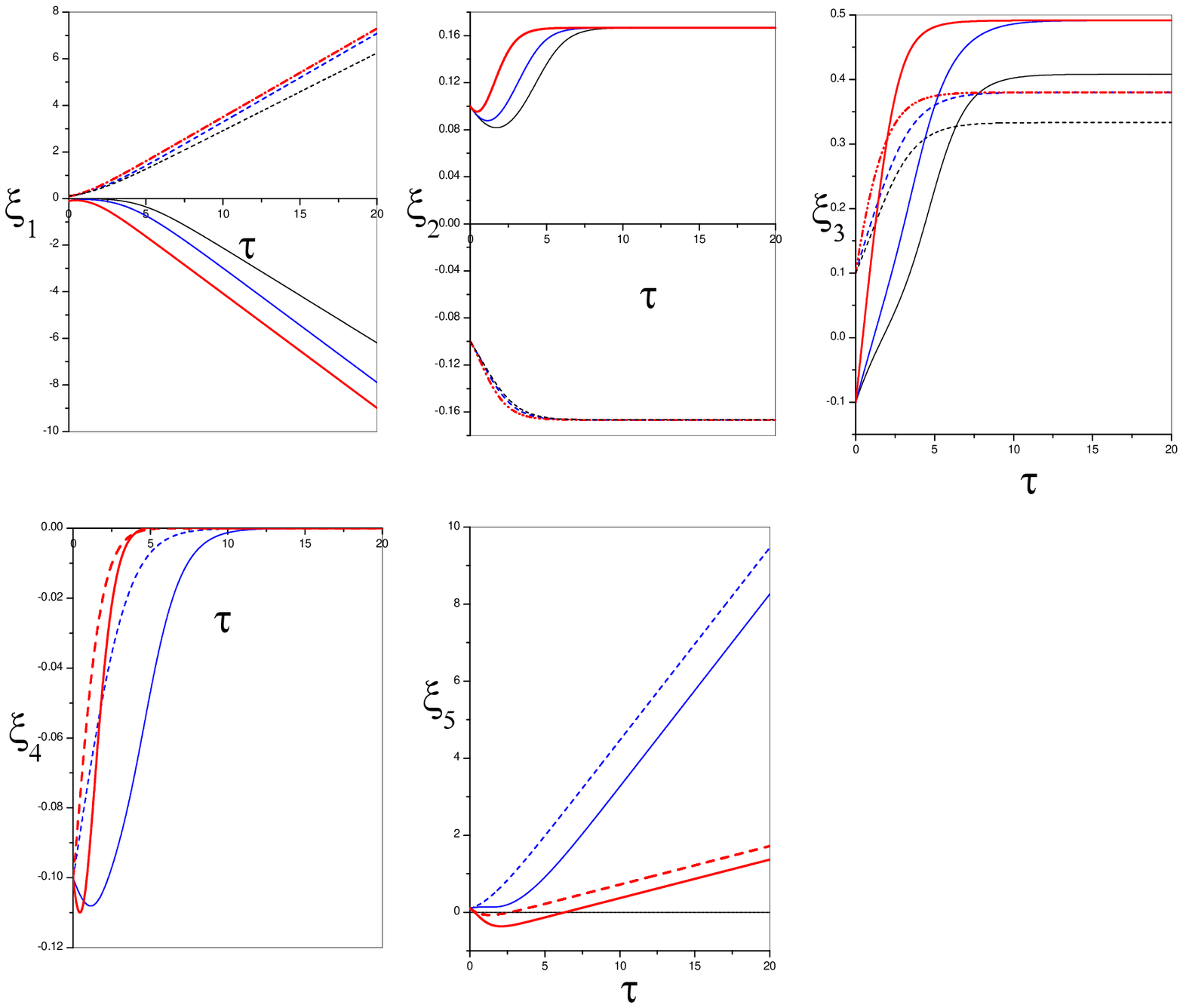}}
\begin{verse}
\vspace{-0.25cm} \caption{Initial values
$r=\frac{\rho_a}{\rho_b}=1.5,\xi_1=-\xi_2=\xi_3=\xi_4=\xi_5=0.1$
for bubble;$-\xi_1=\xi_2=-\xi_3=\xi_4=\xi_5=0.1$ for spike.Plot
showing variation of $\xi_1,\xi_2$,growth rate $\xi_3,\xi_4$ and
transverse displacement $\xi_5$ of bubble and spike with $V_a=
V_b=0.0$ for solid black line-spike and broken black line for
bubble.$V_a=0.1,V_b=0.5$ for broken blue line-bubble and solid
blue for spike,$V_a=0.5,V_b=0.1$,broken red line for bubble and
solid red line-spike.} \label{fig:1}
\end{verse}
\end{figure}

\begin{figure}[p]
\vbox{ \hskip 1.5cm \epsfxsize=12cm \epsfbox{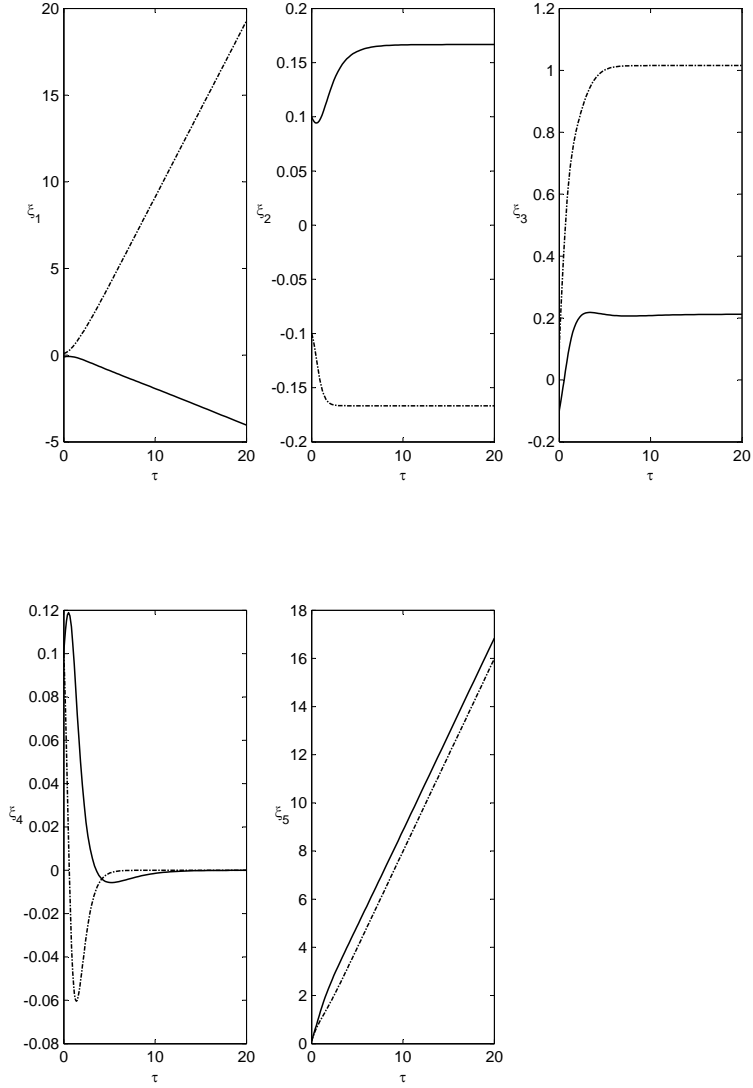}}
\begin{verse}
\vspace{-0.25cm} \caption{r=$\frac{\rho_a}{\rho_b}=0.4$;Lower
fluid denser.Dashed line for spike (heavier fluid pushes into
lighter fluid) and unbroken line for
bubble.$V_a=0.8,V_b=-0.6$.Initial condition as in Fig.1.}and for
following relation$\frac{16(1-r)}{15r}<(V_a-V_b)^2.$ \label{fig:2}
\end{verse}
\end{figure}

\begin{figure}[p]
\vbox{ \hskip 1.5cm \epsfxsize=12cm \epsfbox{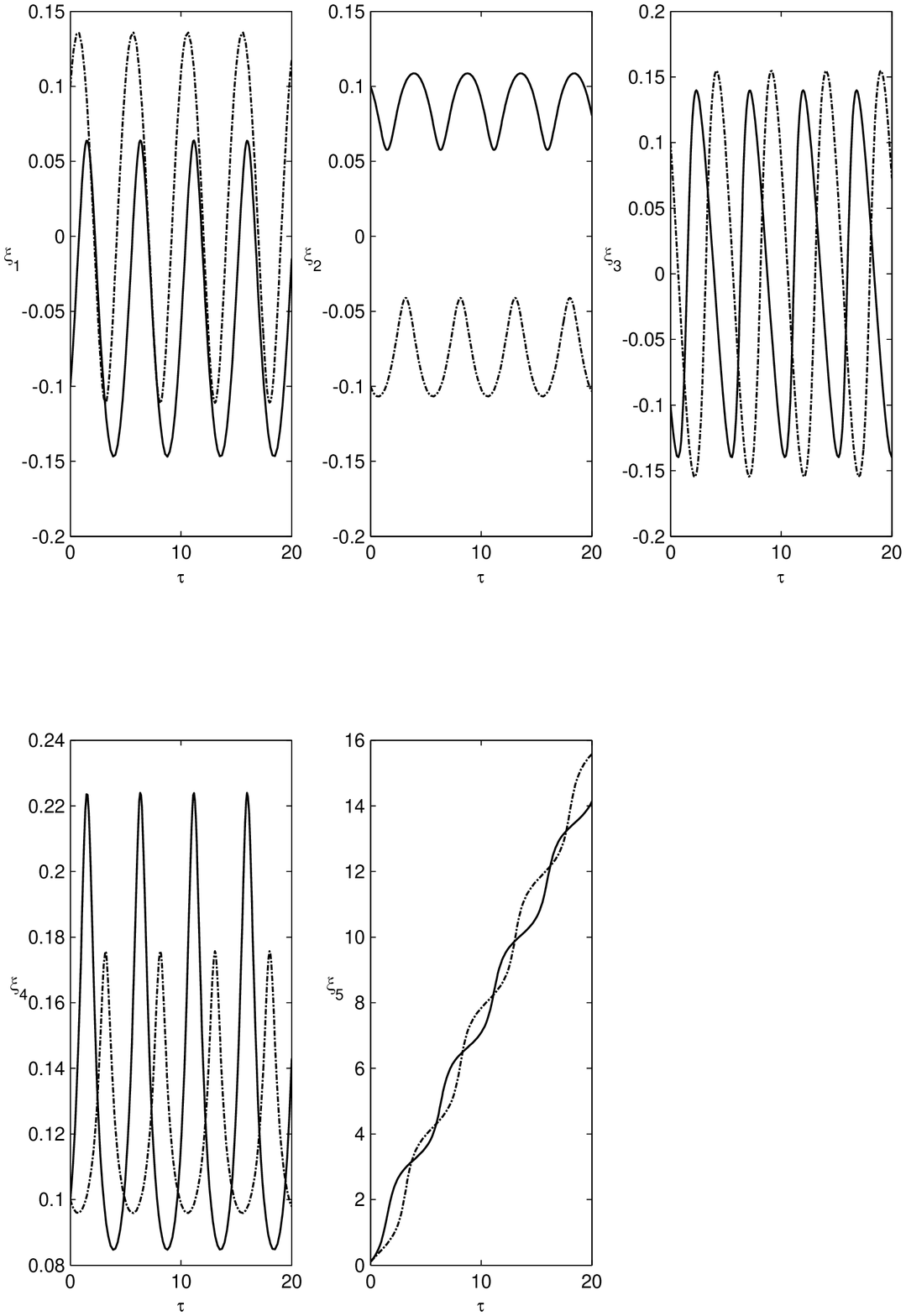}}
\begin{verse}
\vspace{-0.25cm}
\caption{r=0.4,$V_a=0.0,V_b=0.2.(V_a-V_b)^2<\frac{16(1-r)}{15}$
.Initial condition as in Fig.1. Unbroken line for bubble and
dashed line for spike.} \label{fig:3}
\end{verse}
\end{figure}

\begin{figure}[p]
\vbox{ \hskip 1.5cm \epsfxsize=12cm \epsfbox{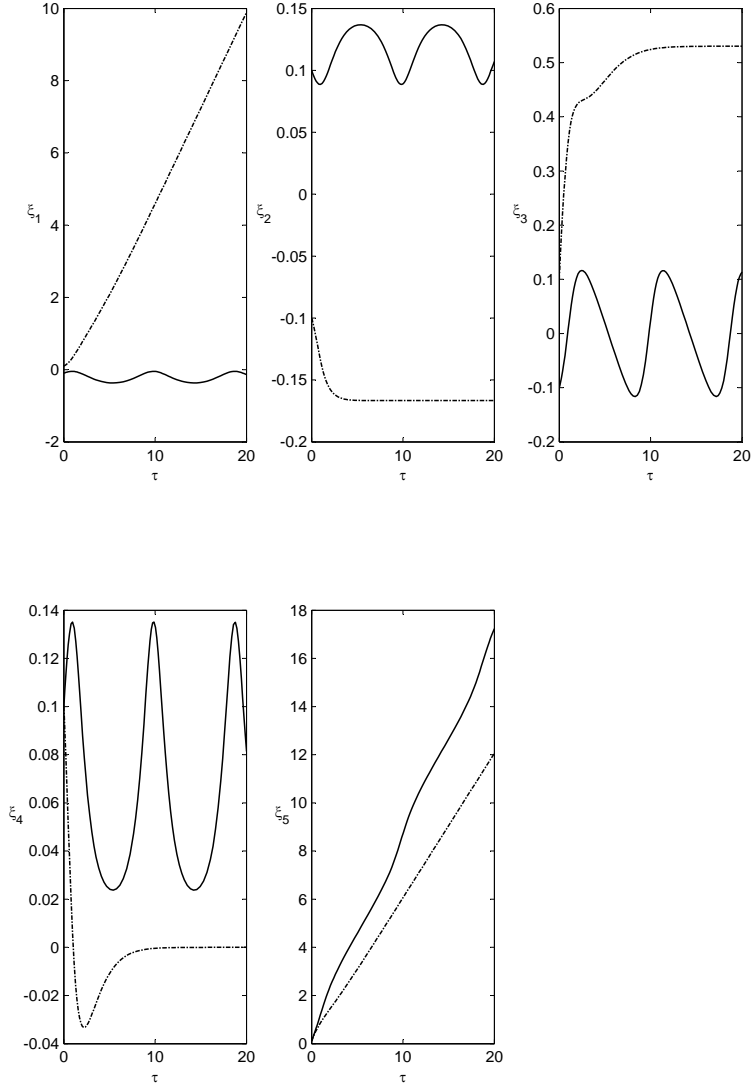}}
\begin{verse}
\vspace{-0.25cm}
\caption{r=0.4,$V_a=0.6,V_b=-0.4;\frac{16(1-r)}{15}<(V_a-V_b)^2<\frac{16(1-r)}{15r}$.Initial
condition as before (Fig.3.).Unbroken line for bubble and dashed
line for spike;height of spike peak increases monotonically with
time (steepening);bubble depth undulates.} \label{fig:4}
\end{verse}
\end{figure}

\begin{figure}[p]
\vbox{ \hskip 1.5cm \epsfxsize=12cm \epsfbox{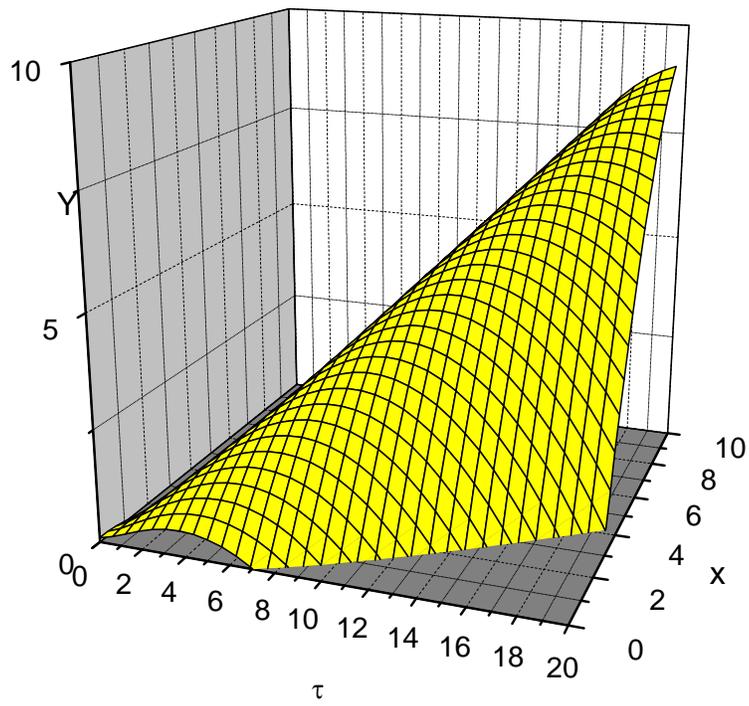}}
\begin{verse}
\vspace{-0.25cm} \caption{3 dimensional plot of spike(Interface
$Y=\eta_0(\tau)+\eta_2(\tau)(x-\eta_1(\tau))^2$)belonging to the
plot given in fig.4.} \label{fig:5}
\end{verse}
\end{figure}

\begin{figure}[p]
\vbox{ \hskip 1.5cm \epsfxsize=12cm \epsfbox{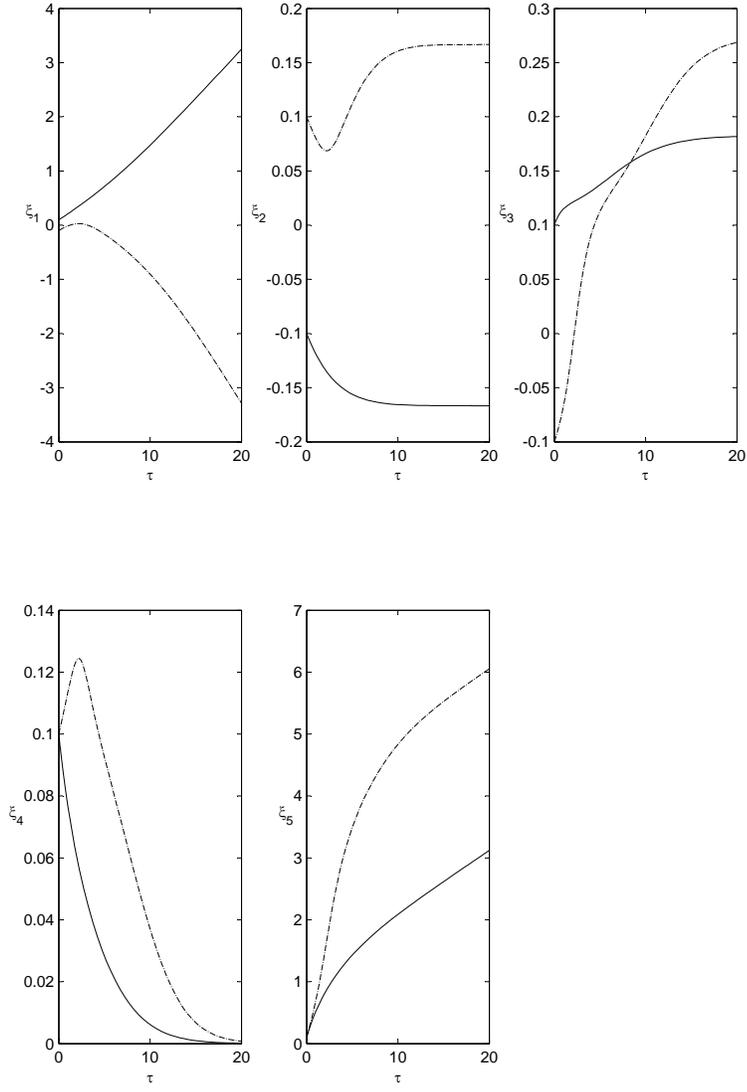}}
\begin{verse}
\vspace{-0.25cm} \caption{Oblique shock: RM and KH instability for
spike (dashed line) and bubble (unbroken line).Initial values as
in fig.1. and $V_a=0.1,V_b=0.5.$} \label{fig:6}
\end{verse}
\end{figure}

\end{document}